# Language of Boolean functions its Grammar and Machine

[1] BIRENDRA KUMAR NAYAK, [2] SUDHAKAR SAHOO

[1] P.G.Department of Mathematics, Utkal University, Bhubaneswar-751004
bknatuu@yahoo.co.uk
[2] Department of CSEA, SIT, Silicon Hills, Patia, Bhubaneswar-751024
sudhakar.sahoo@gmail.com

*Abstract:* -. In this paper an algorithm is designed which generates in-equivalent Boolean functions of any number of variables from the four Boolean functions of single variable. The grammar for such set of Boolean function is provided. The Turing Machine that accepts such set is constructed.

## 1 Introduction

The wide range of applications of Boolean functions in various fields of Computer Science, Mathematics, Engineering etc. and particularly the application of our area of interest like description of Cellular Automata [4, 6, 7], Cryptography [1, 3], image processing [10], Data Compression, Pattern classification, Data mining and so on has raised curiosity to examine the structure of the set of Boolean functions in a more systematic way. As for $n$-variable Boolean functions the number of Boolean functions is $2^{2^n}$ and when $n$ becomes large it is practically infeasible to handle all these Boolean functions even using worlds fastest super computer available today or which will come in the future [3]. In literature as well as in many different papers one can easily find the different ways to represent and analyze the characteristics of Boolean functions. Some of these methods are using the Truth Table, using Boolean Algebra particularly different normal forms like Conjunctive Normal Form (CNF), Disjunctive Normal Form (DNF), Algebraic Normal Form (ANF), using Cellular Automata [4, 6], Matrix Theory [10, 11], Boolean Derivatives [6], State Transition Diagrams (STD's) [4] etc. In this paper the study is made by looking upon the set of all $n$-variable Boolean functions as a formal language [8, 9].

Starting from the Truth Table method which is present in section 2, it is shown in section 3 that a recursive relation exists which can be used to generate any Boolean function of any number of variables from the in equivalent Boolean functions of single variable. With such set of Boolean functions, when assumed as an alphabet a formal grammar is found to exist as shown in section 4 with suitable production rules, which generate any Boolean function of any number of variables in a formal language. Having found that such set of Boolean functions is a formal language, a Turing Machine is constructed in the same section, which accepts this language.

## 2 Boolean functions and their representation

A Boolean function $f(x_1, x_2, ..., x_n)$ on $n$-variables is defined as a mapping from $\{0,1\}^n$ into $\{0,1\}$. It is also interpreted as the output column of its truth table $f$, which is a binary string of length $2^n$. For $n$-variables the number of Boolean functions is $2^{2^n}$ and each Boolean function is denoted as $f_R^n$ known as the function number $R$ (also interpreted as rule number $R$), in $n$-variable. Here $R$ is the decimal equivalent of the binary sequence (starting from bottom to top) of the function in the Truth Table, which obeys the Wolframs naming convention reported in [4].

## Example 2.1 [One variable Boolean function]

In one variable the number of Boolean functions is $2^{2^1} = 4$. These 4 Boolean functions can be arranged following Wolfram convention [4] in a table as shown below.

| $X_1$ | $f_0^1$ | $f_1^1$ | $f_2^1$ | $f_3^1$ |
|---|---|---|---|---|
| 0 | 0 | 1 | 0 | 1 |
| 1 | 0 | 0 | 1 | 1 |

[Table 1]

In 2 variables the number of Boolean functions is $2^{2^2} = 16$ and its truth table is shown below.

| $X_1$ | $X_2$ | $f_0^2$ | $f_1^2$ | $f_2^2$ | $f_3^2$ | $f_4^2$ | $f_5^2$ | $f_6^2$ | $f_7^2$ | $f_8^2$ | $f_9^2$ | $f_{10}^2$ | $f_{11}^2$ | $f_{12}^2$ | $f_{13}^2$ | $f_{14}^2$ | $f_{15}^2$ |
|---|---|---|---|---|---|---|---|---|---|---|---|---|---|---|---|---|---|
| 0 | 0 | 0 | 1 | 0 | 1 | 0 | 1 | 0 | 1 | 0 | 1 | 0 | 1 | 0 | 1 | 0 | 1 |
| 0 | 1 | 0 | 0 | 1 | 1 | 0 | 0 | 1 | 1 | 0 | 0 | 1 | 1 | 0 | 0 | 1 | 1 |
| 1 | 0 | 0 | 0 | 0 | 0 | 1 | 1 | 1 | 1 | 0 | 0 | 0 | 0 | 1 | 1 | 1 | 1 |
| 1 | 1 | 0 | 0 | 0 | 0 | 0 | 0 | 0 | 0 | 1 | 1 | 1 | 1 | 1 | 1 | 1 | 1 |

[Table 2]

Like the two-variable functions, the three variable ones can also be represented in its Truth Table where the number of Boolean functions is $2^{2^3} = 256$. In this way $n$-variable Boolean functions can be defined using the Truth Table method.

## 3 Recursive Algorithm to generate Boolean functions in any number of variables

In Table 2 the first two bit and the last two bit from any 4 bit string of a 2-variable Boolean function is either $\begin{pmatrix}0\\0\end{pmatrix}$ or $\begin{pmatrix}1\\0\end{pmatrix}$ or $\begin{pmatrix}0\\1\end{pmatrix}$ or $\begin{pmatrix}1\\1\end{pmatrix}$ which are basically 1-variable Boolean functions shown in table 1. So any 2-variable Boolean function is the concatenation of any two 1-variable Boolean function $f_0^1$ or $f_1^1$ or $f_2^1$ or $f_3^1$. Similarly 3-variable Boolean functions can be constructed from two 2-variable Boolean functions and it can also be generalized from $n$-variable Boolean functions to get all $n+1$-variable Boolean functions.

The recursive formula to generate $n+1$ variable Boolean functions is as follows:

$$f_0^1 = \begin{pmatrix}0\\0\end{pmatrix}, f_1^1 = \begin{pmatrix}1\\0\end{pmatrix}, f_2^1 = \begin{pmatrix}0\\1\end{pmatrix}, f_3^1 = \begin{pmatrix}1\\1\end{pmatrix},$$

$$f_{l.2^{2^n}+m}^{n+1} = [f_m^n, f_l^n]^T, for\ 0 \leq l, m \leq 2^{2^n} - 1, and\ n \geq 1$$

Where the operations Transposition and Concatenation are defined as $[f_m^n, f_l^n]^T = \begin{pmatrix}f_m^n\\f_l^n\end{pmatrix}$.

**Illustration:** Here is an example to get function number 9 in 2-variable Boolean function from two 1-variable Boolean functions using the above recursive formula.

$$[f_1^1, f_2^1]^T = \begin{pmatrix} f_1^1 \\ f_2^1 \end{pmatrix} = \begin{pmatrix} 1 \\ 0 \\ 0 \\ 1 \end{pmatrix} = f_9^2$$

## 4 Language of Boolean functions

Let $\sum = \{f_0^1, f_1^1, f_2^1, f_3^1\} = \left\{ \begin{pmatrix} 0 \\ 0 \end{pmatrix}, \begin{pmatrix} 1 \\ 0 \end{pmatrix}, \begin{pmatrix} 0 \\ 1 \end{pmatrix}, \begin{pmatrix} 1 \\ 1 \end{pmatrix} \right\}$

$\sum^+$ is the set of all binary strings of even length over the operation transposition and concatenation. Language of Boolean functions L is a subset of $\sum^+$. Next two sections show the grammar and the Turing machine, which describes this language.

### 4.1 Grammar for Boolean functions

The Grammar $G = (V, T, P, S)$, which describes the language of Boolean functions, is as follows:
$G = (V, T, P, S)$
Where
The set of variables $V = \{S, L, R, L_h, [, ]\}$

The set of terminals $T = \{f_0^1, f_1^1, f_2^1, f_3^1\} = \left\{ \begin{pmatrix} 0 \\ 0 \end{pmatrix}, \begin{pmatrix} 1 \\ 0 \end{pmatrix}, \begin{pmatrix} 0 \\ 1 \end{pmatrix}, \begin{pmatrix} 1 \\ 1 \end{pmatrix} \right\}$

$S$ is the starting variable and the set of production rule $P$ consists of 108 productions are as follows:
$P$:
$S \to [R f_0^1] / [R f_1^1] / [R f_2^1] / [R f_3^1] / f_0^1 / f_1^1 / f_2^1 / f_3^1$

$[R f_0^1], [R f_1^1], [R f_2^1], [R f_3^1] \to f_0^1 f_0^1 R / f_0^1 f_1^1 R / f_0^1 f_2^1 R / f_0^1 f_3^1 R / f_1^1 f_0^1 R / f_1^1 f_1^1 R / f_1^1 f_2^1 R / f_1^1 f_3^1 R /$
$\qquad f_2^1 f_0^1 R / f_2^1 f_1^1 R / f_2^1 f_2^1 R / f_2^1 f_3^1 R / f_3^1 f_0^1 R / f_3^1 f_1^1 R / f_3^1 f_2^1 R / f_3^1 f_3^1 R$

$f_3^1 L, f_2^1 L, f_1^1 L, f_0^1 L \to L f_0^1 / L f_1^1 / L f_2^1 / L f_3^1$

$f_3^1 L_h, f_2^1 L_h, f_1^1 L_h, f_0^1 L_h \to L_h f_0^1 / L_h f_1^1 / L_h f_2^1 / L_h f_3^1$

$R] \to L] / L_h$

$[L \to [R$

$[L_h \to \in$

**Illustration:**

$S \Rightarrow [R f_0^1] \Rightarrow [f_0^1 f_3^1 R] \Rightarrow [f_0^1 f_3^1 L_h \Rightarrow [f_0^1 L_h f_3^1 \Rightarrow [L_h f_0^1 f_3^1 \Rightarrow \in f_0^1 f_3^1 \Rightarrow f_0^1 f_3^1 = \begin{pmatrix} 0 \\ 0 \\ 1 \\ 1 \end{pmatrix} = f_{12}^2$

## 4.2 Turing Machine for Boolean functions in n-variable

The TM, which accepts the above language, is constructed as follows:

$M = (Q, \Sigma, \Gamma, \delta, q_1, q_{accept}, q_{reject})$

Where

The set of states $Q = \{q_1, q_2, q_3, q_4, q_5, q_{accept}, q_{reject}\}$,

Alphabet $\Sigma = \{f_0^1, f_1^1, f_2^1, f_3^1\} = \left\{\begin{pmatrix}0\\0\end{pmatrix}, \begin{pmatrix}1\\0\end{pmatrix}, \begin{pmatrix}0\\1\end{pmatrix}, \begin{pmatrix}1\\1\end{pmatrix}\right\}$,

The set of writing symbols $\Gamma = \{f_0^1, f_1^1, f_2^1, f_3^1, x, B\}$,

The initial state is $q_1$ and

The transition function $\delta$ is defined as follows:

$\delta(q_1, f_i^1) = (q_2, B, R)$ for $i = 0, 1, 2, 3$

$\delta(q_2, B) = (q_{accept}, B, R)$

$\delta(q_2, x) = (q_2, x, R)$

$\delta(q_2, f_i^1) = (q_3, x, R)$ for $i = 0, 1, 2, 3$

$\delta(q_3, x) = (q_3, x, R)$

$\delta(q_3, f_i^1) = (q_4, f_i^1, R)$ for $i = 0, 1, 2, 3$

$\delta(q_3, B) = (q_5, B, L)$

$\delta(q_4, x) = (q_4, x, R)$

$\delta(q_4, f_i^1) = (q_3, x, R)$ for $i = 0, 1, 2, 3$

$\delta(q_4, B) = (q_{reject}, B, R)$

$\delta(q_5, f_i^1) = (q_5, f_i^1, L)$ for $i = 0, 1, 2, 3$

$\delta(q_5, x) = (q_5, x, L)$

$\delta(q_5, B) = (q_2, B, R)$

$\delta(q_1, B) = (q_{reject}, B, R)$

$\delta(q_1, x) = (q_{reject}, x, R)$

**Illustration:**

(i) $q_1 f_0^1 f_1^1 f_2^1 \to B q_2 f_1^1 f_2^1 \to B x q_3 f_2^1 \to B x f_2^1 B q_{reject}$.

(ii) $q_1 f_0^1 f_1^1 \to B q_2 f_1^1 \to B x q_3 B \to B q_5 x \to q_5 B x \to B q_2 x \to B x q_2 B \to B x B q_{accept}$.

## 5. Conclusion

In this paper it is shown that any n-variable Boolean functions can be treated as a language, which can be, derived recursively from one variable Boolean function. Also the grammar, which derives this language and the machine, which accepts it, is found out. We are investigating into the production rules which will produce Boolean functions with specific characteristics like degree of Non-linearity, Balanced ness etc.

*References:*

[1] Yuriy Tarannikov, New constructions of Resilient Boolean Functions with maximum non-linearity, Mech. and Math. Department Moscow states University 119899 Moscow, Russia.

[2] Subhamoy Maitra and Enes pasalic, Further constructions of Resilient Boolean Functions with very high non-linearity.

[3] Claude Carlet, Boolean functions for Cryptography and Error correcting codes.

[4] S. Wolfram, Statistical mechanics of Cellular Automata, *Rev Mod Phys.* 55,601-644 (July 1983)

[5] E. Filion, C. Fontaine, High Non-linear Balanced Boolean Functions with good correlation immunity, Advanced in cryptography, Eurocrypt'98, Helsinki, Finland, lecture notes in Comp.Sc., Vol 1403, 1998, pp.475-488

[6] G. Y. Vichniac, Boolean Derivatives On Cellular Automata, Physica D (Nonlinear Phenomena),volume 45,1990, pp-63-74.

[7] N.H. Packard and S.WolForm, Two-dimensional cellular automata, *Journal of Statistical Physics*, 38 (5/6) 901-946, (1985).

[8] Problem solving in Automata, Languages and Complexity by Ding-zhu-Cu and ker I ko, John Wiley and Sons Inc.(2001).

[9] Introduction to the Theory of Computation, by Michael Sipser, International student edition.

[10] P.P.Choudhury, B.K. Nayak, S. Sahoo, *Efficient Modelling of some Fundamental Image Transformations*. Tech.Report No. ASD/2005/4, 13 May 2005.

[11] P.P. Choudhury, K. Dihidar, Matrix Algebraic formulae concerning some special rules of two-dimensional Cellular Automata, International journal on Information Sciences, Elsevier publication, Volume 165, Issue 1-2.